\documentclass[reprint, twocolumn, prl, superscriptaddress, amsmath,amssymb, aps,]{revtex4-1}

\usepackage{graphicx}
\usepackage{array}
\usepackage{dcolumn}
\usepackage{bm}
\usepackage{listings}
\usepackage{scalefnt}
\usepackage{xcolor}
\usepackage{booktabs}
\usepackage{amsthm,amsfonts}

\usepackage{braket}

\begin{document}

\title{Dynamical Casimir Effect for Gaussian Boson Sampling}

\author{Borja Peropadre}
\address{Quantum Information Processing group, Raytheon BBN Technologies, 10 Moulton Street, Cambridge, Massachusetts 02138, USA}

\author{Joonsuk Huh}
\address{Department of Chemistry, Sungkyunkwan University, Suwon 440-746, Korea}

\author{Carlos Sab\'in}
\email{Correspondence to:  csl@iff.csic.es}
\address{Instituto de F\'isica Fundamental, CSIC, Serrano 113-bis, 28006 Madrid, Spain}

\date{\today}
             
\begin{abstract}
We show that the Dynamical Casimir Effect (DCE), realized on two multimode coplanar waveguide resonators, implements a gaussian boson sampler (GBS). The appropriate choice of the mirror acceleration that couples both resonators translates into the desired initial gaussian state and many-boson interference in a boson sampling network. In particular, we show that the proposed quantum simulator naturally performs a classically hard task, known as scattershot boson sampling.  Our result unveils an unprecedented computational power of DCE, and paves the way for using DCE as a resource for quantum simulation.   

\end{abstract}

\maketitle

\section*{Introduction} 

The Dynamical Casimir Effect (DCE) consists in the generation of photons out of the vacuum of a quantum field by means of the abrupt modulation of boundary conditions -e.g. a mirror oscillating at speeds comparable to the speed of light. Predicted in 1970 \cite{moore}, an experimental demonstration remained elusive until 2011, when it was implemented in a superconducting circuit architecture \cite{casimirwilson}. In addition to its fundamental interest, it has been shown that the radiation generated in the DCE displays entanglement and other forms of quantum correlations \cite{nonclassicaldce,discord,ipsteering,casimirsimone}, which poses the question of its utility as a resource for the heralded quantum technological revolution. As an example, small-scale continuous variable cluster states of four electromagnetic field modes have been shown to be in principle possible \cite{cluster}. While this represents a preliminary step for a continuous variable one-way quantum computer, its scalability has not yet been demonstrated, hence the usefulness of DCE for quantum computing tasks remains unclear.

\begin{figure}[t!]
\begin{center}
\includegraphics[width=\linewidth]{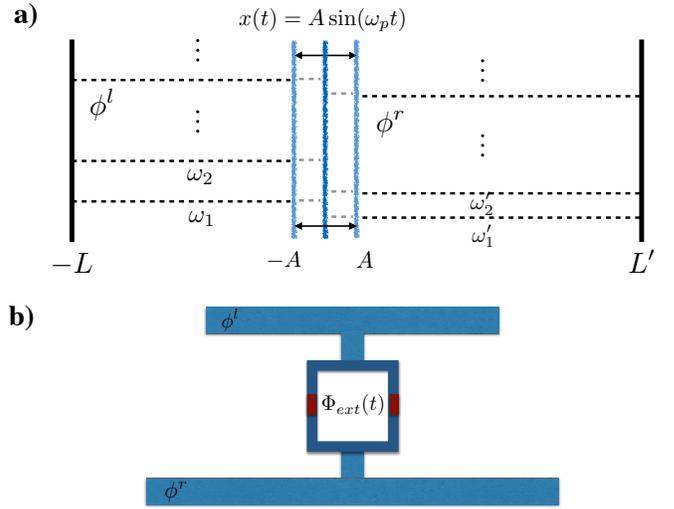}
\caption{a) Two resonators of different lengths $L$ and $L'$ -and thus different energy spectra $\{\omega_{l}\}$, $\{\omega'_{l}\}$ for the corresponding field modes $\phi^l$ and $\phi^r$- sharing a common wall which moves harmonically with amplitude $A$ and frequency $\omega_p$. b) Two superconducting transmission line resonators coupled through a dc-SQUID's acting as a tunable common mirror. The modulation of the external magnetic flux $\Phi_{ext}(t)$ amounts to an effective motion of the mirror.
}\label{fig:name1}
\end{center}
\end{figure}

In this work, we establish a bridge between multimode parametric amplification induced by the modulation of boundary conditions - for which DCE is a paradigmatic case- and a non-universal quantum computing device, known as boson sampling (BS) \cite{Aaronson2011}. 

BS has recently gained a great deal of attention, as it solves a tailor-made problem --the problem of sampling from the output distribution of photons in a linear-optics network-- that is widely believed to be intractable in any classical device. Thus it represents a promising avenue for proving the long-sought quantum supremacy \cite{preskill12}. We consider Gaussian BS (GBS) and in particular scattershot BS, a generalization of the original BS problem which is known to be equivalent in terms of computational complexity \cite{Lund2014}.  GBS have been proven to be of practical interest in reconstructing the Franck-Condon profile -a central problem in molecular spectroscopy,- both theoretically \cite{Huh2014} and in a recent experimental trapped-ion implementation \cite{Huh2017}.   We show that GBS can be implemented in a  superconducting circuit architecture by exploiting the possibility of multimode parametric amplification by means of the modulation of boundary conditions. We propose a setup consisting of two superconducting resonators coupled through a superconducting quantum interferometric device (SQUID) \cite{Peropadre13} (see Fig. \ref{fig:name1}). The resonators possess different lengths and thus different energy spectra and the SQUID plays the role of a shared tunable mirror-like boundary condition. The modulation of the external magnetic flux threading the SQUID implements an effective motion of the mirror whose corresponding Bogoliubov transformation results in multimode parametric amplification. We show that suitable choices of the SQUID pumping are able to implement the operations of a GBS -namely two-mode squeezers, beam-splitters and phase shifters.  In this way, we show how the DCE can be exploited as a quantum simulator of GBS.  Moreover, we will discuss how DCE is by itself a physical effect that is hard to simulate on a classical computer.  

\section*{Methods}

The DCE was observed in an open microwave coplanar waveguide interrupted by a single SQUID operated well below its plasma frequency \cite{casimirwilson} . Under the latter condition, the SQUID implements an effective mirror-like boundary condition. Ultrafast variation of the magnetic flux threading the SQUID amounts to motion of the mirror at relativistic speeds, which generates a two-mode squeezing operation on the microwave field propagating along the transmission line. In particular, for an initial vacuum field state the modulation of the boundary condition results in generation of pairs of photons, a process which is resonantly enhanced if the mirror moves at a frequency matching the sum of the frequencies of the emitted photons.

Obviously, the DCE can be produced as well for different boundary conditions, such as the ones of a superconducting resonator interrupted by one \cite{simoen} or two \cite{idathesis} SQUIDs. Moreover, we can think of the DCE as a particular instance of multimode parametric amplification induced by the modulation of boundary conditions, as we shall see in the following.

Let us consider a one-dimensional (1D) superconducting resonator in the presence of one or two movable boundary conditions. In the absence of any flux modulation, the resonator field $\phi$ is characterised by a set of creation and annihilation operators $\{a_l,a^{\dagger}_l\}$ associated to the set of solutions $\{u_l, u_l^*\}$ of the 1D massless Klein-Gordon wave equation --plane waves -- with the corresponding boundary conditions -- e.g. the well-known standing waves in the case of a perfect resonator\cite{Blais04}.

The modulation of the SQUIDs  changes the boundary conditions of the field, generating a new set of solutions $\{v_l,v_l^*\}$ and the corresponding new set of operators $\{a'_l,a'^{\dagger}_l\}$. Both sets are related by means of a Bogoliubov transformation:
\begin{equation}
a'^{\dagger}_l=\sum_j -\beta_{jl}a_{j}+\alpha_{jl}a^{\dagger}_j.
\label{eq:bogomtrix}
\end{equation}
where the Bogoliubov coefficients $\{\alpha_{jl},\beta_{jl}\}$ are given by the inner product:
\begin{eqnarray}
\alpha_{jl}&=&(v_j, u_{l})\nonumber\\
\beta_{jl}&=&-(v_j,u_{l}^*).
\end{eqnarray}
Therefore, they depend on the particular initial boundary conditions -which determine the functions $u_l$- and the particular type of boundary modulation -which determine the functions $v_l$. In the case of small boundary oscillations characterised by a dimensionless amplitude $\delta$, the Bogoliubov coefficients can be computed perturbatively in a variety of a cases including single- \cite{singleDCE} and two-wall oscillations \cite{doubleDCE}. More general continuous motion of the two walls, such as the one required to mimic the motion of an accelerated cavity which is rigid in its proper frame, can also be addressed perturbatively \cite{louko}. In all these cases, the Bogoliubov coefficients depend on the features of the boundary modulation, for instance, the number of external pumps, together with their corresponding frequencies and durations.

Notice that the set $\{\alpha_{jl}\}$ characterizes phase-shifting ($j=l$) and  beam-splitting ($j\neq l$) among the modes \cite{zakka}, while $\{\beta_{jl}\}$ generates two-mode squeezing \cite{casimirwilson}. Therefore, we conclude that the modulation of the boundary conditions of a superconducting resonator is equivalent to a multimode parametric amplifier consisting of a set of tunable phase shifters, beam splitters and two-mode squeezers, which can be adjusted by suitably selecting the number, frequency and duration of external pumps.

A remarkable example is the DCE, where a modulation of frequency $\omega_p$ generates Bogoliubov coefficients $\beta_{lj}$ growing linearly in time only for the modes in a resonance condition $\omega_p=\omega_l+\omega_j$ -all the other Bogoliubov coefficients being negligible. Similarly, another resonance frequency $\omega_p=|\omega_l-\omega_j|$ would make the corresponding $\alpha_{lj}$ to increase linearly in time \cite{louko}. 
Clearly, a series of operations of this kind with suitable frequencies and duration times can generate a desired combination of beam splitters and two-mode squeezers.

\section*{Results}

\begin{figure}[t!]
\begin{center}
\includegraphics[width=\linewidth]{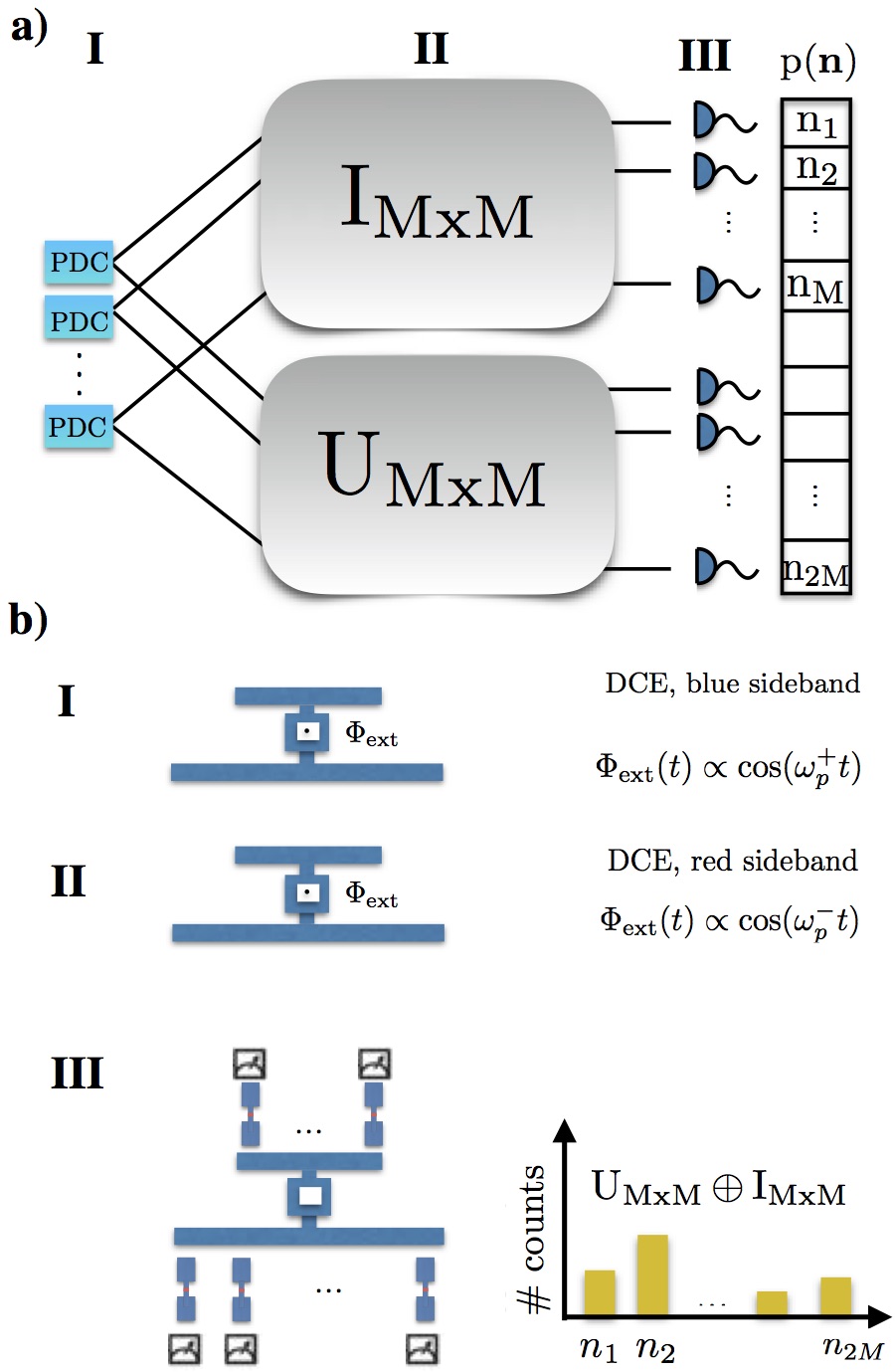}
\caption{a) Schematics of a gaussian boson sampler (GBS). Parametric down conversion sources (PDCs) generate two-mode squeezed states (I), half of them undergo idle evolution whereas the other half evolve under the unitary $U$ (II). Single photon detectors collect the statistics $p(n)$ at the output of the photonic network (III) b) GBS using a DCE-like dynamics in multimode superconducting resonators. The distinct modulation of the external magnetic flux $\Phi_{ext}(t)$ amounts to blue (I) or red (II) sidebands that implements both initial state and unitary dynamics. Finally, ancilla qubits (III) could be employed to perform a Ramsey measurement scheme in order to resolve $P(n)$}.
\label{fig:name}
\end{center}
\end{figure}

In the following, we will show how to use this scheme to implement a GBS protocol (see Fig.~\ref{fig:name}a)~\cite{Lund2014}. 

An important problem is the well known lack of addressability in a harmonic oscillator.  We need to generate beam splitters interactions only between a selected pair of modes, without further driving unwanted transition between modes.  A way to overcome this problem is by considering a pair of resonators of different lengths - and hence different energy spectrum- confining two resonator fields $\phi^r$, $\phi^l$ and sharing a tunable mirror (see Fig. \ref{fig:name1}a). In particular, to make sure this addressability condition holds for any pair of modes, it is convenient to choose resonators of incommensurate lengths.This condition can be relaxed, if the we set the appropriate cutoff in the energy spectrum of the resonators.

Therefore the collection of modes $u$ consists of two collections of modes $u^l,u^r$ given by the solutions to the Klein-Gordon 1D equation - plane waves- after imposing Dirichlet boundary conditions at points, say $-L$, $0$ for $u^l$ and $0$, $L'$ for $u^r$:
\begin{eqnarray}
u^l_l(x,t)&=&\frac{1}{\sqrt{\pi\,l}} \operatorname{sin}(k_l\,(x+L))\,e^{-i\,\omega_l t}\nonumber\\
u^r_l(x,t)&=&\frac{1}{\sqrt{\pi\,l}} \operatorname{sin}(k'_l\,x)\,e^{-i\,\omega'_l t}
\end{eqnarray}
where $\omega_{l}=v\,k_l=\frac{l\pi\,v}{L}$, $\omega'_l=v\,k'_l=\frac{l\pi\,v}{L'}$ and $v$ is the propagation speed of the field. 

The collection of modes $v$ is given by the transformed solutions of the field confined in both resonators when the common mirror -initially placed at $x=0$- undergoes an effective harmonic motion of frequency $\omega_p$ and amplitude $A$, given by $x(t)=A\sin{\omega_p\,t}=\delta L'\sin{\omega_p\,t}$, where $\delta=A/L'\ll 1$ is a dimensionless small parameter. Then we can obtain the Bogoliubov coefficients as a perturbative expansion in $\delta$. In particular, the first order of the expansion will be given by \cite{louko}:
\begin{eqnarray}
\alpha_{jl}^{(1)}(t)&=&i\,(\omega_j'-\omega_{l}){}_0\alpha_{jl}^{(1)}\int^{t}_{0}dt' e^{-i(\omega_j'-\omega_{l})t'}\sin{\omega_p t'}\nonumber\\
\beta_{jl}^{(1)}(t)&=&i\,(\omega_j'+\omega_{l}){}_0\beta_{jl}^{(1)}\int^{t}_{0}dt' e^{-i(\omega_j'+\omega_{l})t'}\sin{\omega_p t'},\label{eq:bogost}
\end{eqnarray}
where $\omega_{l}$,$\omega_j'$ are the frequencies of the modes $u_{l}$ and $v_j$ respectively and $_0\alpha_{jl}^{(1)}$, ${}_0\beta_{jl}^{(1)}$ are the Bogoliubov coefficients associated to the transformation induced by a single change of mirror position $x=\delta L'$ at $t=0$.
Then:
\begin{eqnarray}\label{eq:bogos}
_0\alpha_{jl}&=&-\frac{i}{c}\int_{-L}^{L'} dx (v_j\partial_t\,u_{l}^*-u_{l}^*\partial_t\,v_j)|_{t=0}\nonumber\\
_0\beta_{jl}&=&\frac{i}{c}\int_{-L}^{L'} dx (v_j\partial_t\,u_{l}-u_{l}\partial_t\,v_j)|_{t=0}.
\end{eqnarray}
Now, if we select a mode $u^l$ in the left cavity and a mode $v^r$ in the right cavity:
\begin{equation}
v^r_j(x,t)=\frac{1}{\sqrt{\pi\,j}} \operatorname{sin}\left (\frac{k'_j}{1-\delta}\,(x-\delta L' )\right )\,e^{\frac{-i\,\omega'_j t}{1-\delta}},
\end{equation}
we find that:
\begin{eqnarray}\label{eq:bogospart}
_0\alpha_{jl}&=&_0\alpha_{jl}^{(1)}=-\frac{(-1)^l j\,l\,\pi^3A^2}{6L}\left (\frac{l}{L}-\frac{j}{L'}\right )\delta\nonumber\\
_0\beta_{jl}&=&_0\beta_{jl}^{(1)}=-\frac{(-1)^l j\,l\,\pi^3A^2}{6L}\left (\frac{l}{L}+\frac{j}{L'}\right )\delta.
\end{eqnarray}

By inspection of Eq. (\ref{eq:bogost}),we see that Bogoliubov coefficients oscillate in time, unless the pumping frequency $\omega_p$ is either 
\begin{equation}
\omega_p^-=|\omega_j'-\omega_{l}|=\left |\pi\,c\left (\frac{j}{L'}-\frac{l}{L}\right )\right |
\end{equation}
or
\begin{equation}
\omega_p^+=\omega_j'+\omega_{l}= \pi\,c\left (\frac{j}{L'}+\frac{l}{L}\right ).
\end{equation}
In the former case the corresponding $\alpha_{jl}$ contain a term that grows monotonically in time, while in the latter the same happens for $\beta_{jl}$. In both alternative scenarios, after a time $\omega_p^\pm\,t\gg 1$, we can neglect all the oscillations, so we can assume that only the resonant terms are non-zero. In the first case (by simplicity, we assume $L'>L$):
\begin{equation}\label{eq:bogospartexp}
\alpha_{jl}=\frac{(-1)^l\,\omega_j'\omega_{l}\omega_p^2\,A^3}{12\,c^3} t, 
\end{equation}
while in the second one we obtain exactly the same expression for $\beta_{jl}$, with a minus sign on the front.
 
Notice that in our setup, the difference or sum of two mode frequencies when the modes correspond to different resonators is different for each pair. 
Thus Eq. (\ref{eq:bogospartexp}) and the corresponding expression for the $\beta_{jl}$ illustrate the ability of implementing a beam splitter or a two-mode squeezer between a pair of selected modes by means of a suitable choice of the pumping frequency $\omega_p$. Then, the magnitude of the Bogoliubov coefficients can be adjusted with the choice of the amplitude $A$ and duration $t$ of the oscillation. In order to obtain random coefficients, we can randomize the value of $A$ by means of a random number generator. Note that, while  obviously $\alpha_{jl}$ and $\beta_{jl}$ cannot be complex with the approach above -- although additional random phases could be added by means of a rotation of the pump \cite{casimirwilson}--,it has been shown that arbitrary random real matrices are  still classically hard to sample, as long as the entries are both positive and negative numbers \cite{Aaronson2011,negative}.
\subsection* {DCE for Gaussian Boson Sampling. Circuit QED implementation}
\subsubsection*{Gaussian Boson Sampling and scattershot boson sampling} 

GBS \cite{Lund2014} is a modification of the original BS problem, where the initial state is Gaussian, as opposed to the initial state of the original BS, which is a Fock state. An example of Gaussian initial state would be a product of several two-mode squeezed states. In particular, we can think of a setup in which half of the output of $n^2$ two-mode squeezers are input into a linear network of $n^2$ optical modes, while the other half is sent directly to single photon detectors. Then, $n$ single photons are detected in the latter half. As shown in \cite{Lund2014} this device is able to solve a randomized version of BS known as scattershot BS, which possesses similar computational complexity as the original problem \cite{Aaronson2011}, and therefore is widely believed to be out of reach classically in the same limit, namely approximately $20$ single photons and $400$ modes. Therefore, the necessary optical operators are: I) two-mode squeezers for the preparation of the initial state, II), beam splitters and phase shifters for the implementation of the linear network \cite{Reck94} and III), single photon detectors.

\subsubsection*{Circuit QED implementation}

The results of the previous section suggest that superconducting circuits are particularly suitable for implementing the above operators. In particular, we propose multimode coupled resonators as laid out in Fig. 1b, to prepare two-mode squeezed states between arbitrary modes $i$, $j$, and realize beam-splitting operations among the modes. This is done by time-evolving the coupled resonators under circuit Hamiltonian
\begin{eqnarray}
H&=&\sum_{k=1}^M \frac{{q_k}^2}{2c}+\frac{1}{2l}\omega_k^2{\phi_k}^2+\sum_{k=1}^M \frac{{q'_k}^2}{2c'}+\frac{1}{2l'}{\omega'_k}^2{\phi'_k}^2\nonumber\\
&-&E_{J,-}\cos\left({{\varphi_{-}}}\right)-E_{J,+}\cos\left({{\varphi_{+}}}\right).
\label{eq:circuit_Ham2}
\end{eqnarray}
The charges $q_k$ are canonical conjugate variables of the flux modes $\phi_k$, which have frequencies $\omega_k=k\pi v/L$, where $L$ is the size of the resonator and $v=1/\sqrt{lc}$ the speed of light in the coplanar waveguide- $l$ and $c$ are the inductance and capacitance per unit length, respectively. Naturally, a similar expression holds for modes $\phi _k '$. $E_{J{\pm}}$ is the Josephson energy of the junctions in the SQUID loop, and $\varphi_{\pm}$ represents the superconducting phase drop across each junction. The second line of Eq.~(\ref{eq:circuit_Ham2}) accounts for the nonlinear inductive energy of the SQUID, and is responsible of the coupling between resonators. This is obvious after imposing the fluxoid quantization relation $\varphi_-=2\pi(\phi'-\phi)/\Phi_0$, $\varphi_{+}=2\pi\Phi_{\text{ext}}/\Phi_0$, being $\Phi_0$ the flux quantum. Introducing the boson creation and annihilation operators, $a_k$, $a^\dagger_k$, and assuming small phase slips $\varphi_{-}$ across the SQUID, the circuit Hamiltonian can be written in the interaction picture as
\begin{equation}
H=-2E_{J}\sum_{j,l}\cos\left({{2\pi\Phi_{\text{ext}}/\Phi_0}}\right)(a_l+a_l^\dagger)(a'_j+{a'}_j^\dagger).
\label{eq:circuit_Ham}
\end{equation}
An external magnetic field through the SQUID $\Phi_{\text{ext}}=\Phi_{lj}^+\cos(\omega^{+}_{p} t+\phi_{lj})$, with $\omega_p^+$ given by (9), yields the effective Hamiltonian $H_{\text{eff}}=\xi_{lj} a^{'\dagger}_la^{\dagger}_j+\text{H.c.}$ that implements a two-mode squeezing operation between modes $i$ and $j$, where the squeezing coefficient $\xi_{lj}\propto{J_1(2\pi\Phi_{lj}^{+}}/\Phi_0)$ --where $J_1$ is the Bessel function of the first kind-- results from Jacobi-Anger expansion, which is indeed the Bogoliubov coefficient $\beta_{lj}$ given by equation (\ref{eq:bogospartexp}). The sequential evolution under (13) for distinct pair of modes $i,j$ and the different frequencies $\omega_p^+$ results in the preparation of the desired initial state $\ket\psi=S^\dagger \ket{\psi_0}$, where $\ket{\psi_0}=\ket{0}^{\otimes 2M}$ , and the squeezing operator $S=\exp{(iHt)}$ results from the time evolution under the Hamiltonian \cite{PhysRevA.95.032327}.

Similarly, pumping the SQUID with a field $\Phi_{\text{ext}}=\Phi_{lj}^-\cos(\omega^{-}_{p} t+\phi_{lj})$ at the frequency $\omega_p^{-}$ given by (8) results in a red sideband Hamiltonian $H_{\text{eff}}=g_{lj} e^{i\phi_{lj}} a^{'\dagger}_la_j+ \text{H.c.}$  that implements beam splitting and phase shifting operations between modes $l,j$. Note that arbitrary relative phase shifts on each mode are implemented either by a period of free evolution -no pumps-, since all the resonator modes possess a different frequency, or by phase-shifting the pump itself with an external phase $\phi_{lj}$. The coupling coefficient $g_{lj}\propto{J_1(2\pi\Phi_{lj}^{-}}/\Phi_0)$ is indeed the Bogoliubov coefficient $\alpha_{lj}$ in equation (\ref{eq:bogospart}).
As shown by Reck et al. \cite{Reck94}, any unitary operator $R_U$ can be decomposed in these passive linear operations, $R_{U}=U^K\otimes\cdots U^1$, where $K=\mathcal{O}(M^2)$, and $U^k$ connects nearest neighbor modes. Interestingly, the circuit depth can be reduced to $K=\mathcal{O} (M\log(M))$ by implementing non-local beam-splitters \cite{Aaronson2011}. This can be naturally implemented in our proposal, as we are dealing with beam splitters in frequency space and then we can connect any pair of modes ${i,j}$ by choosing the right frequency for the pump $|\omega_i-\omega_j|$. This has the advantage of reducing the number of operations considerably.

The final step of the protocol, after evolving under red and blue sideband Hamiltonians, is reading out the number of photons on each mode of the resonators. In the diluted limit of boson sampling, where the number of modes $M=N^2$, more than one excitation per mode is unlikely to occur, for which parity measurements would be enough. In this case, one can perform Ramsey-type measurements, where ancilla qubits are coupled to a different mode of the resonator.  For each qubit $j$ of frequency $\Omega_j$, one applies two $\pi/2$ pulses separated by a time $t_{n,j}=\pi g^2_n/\Delta_{n,j}$, where $\Delta_{n,j}=\omega_n-\Omega_j$ is the qubit-nth cavity mode detuning. With the qubits initially in the ground state, the Ramsey-type measurement maps the even (odd) parity onto the excited (ground) state of the qubit. The qubit state can be finally measured by a projective measurement, revealing the n-th mode state partity, and thus whether there is 0 or 1 photon in such mode \cite{Ramsey}. 
Alternatively, one could also perform number-resolving measurments on each mode of the resonator by measuring a  photon number-dependent energy splitting on the ancillary qubits, as described elsewhere \cite{Peropadre2015microwave}.

We would like to highlight that a similar system of coupled resonators with tunable coupling has already been implemented in the laboratory \cite{coupled}. Using a SQUID as a coupler would not increase the experimental requirements \cite{casimirsimone,Peropadre13}. Multimode parametric amplification by means of SQUID boundary condition modulation has already been reported as well \cite{simoen}. Note that the experimental state -of-the-art in boson sampling with optical setups is still within the regime of low number of photons (three photons up to six modes)\cite{Broome794,Spring798,2013NaPho...7..540T,2013NaPho...7..545C}, while integrated photonic circuits have achieved a maximum number of three photons in thirteen modes probabilistically generated from six SPDC sources \cite{Bentivegnae1400255}. Using this number as benchmark, our setup would require addressing 4 modes per resonator, which is completely within experimental reach in multimode circuit QED setups \cite{PhysRevLett.114.080501}. While there has no been experiments implementing boson sampling with superconducting circuits so far, at least another realistic proposal exists \cite{Peropadre2015microwave}. Our setup might be less resource-consuming, since it only involves two resonators instead of a large array. 

In order to remain within the employed perturbative approximations, we should have $|\beta_{jl}|<0.1$ for all the relevant $j,l$. Using Eq.(\ref{eq:bogospartexp}) with the realistic parameters $A\simeq 1\operatorname{mm}$ and $\omega_p\simeq 10 \operatorname{GHz}$, this implies a feasible time duration of the pulses of around $100 \operatorname{ns}$, for each pulse involved in both the state preparation and the unitary evolution. Since the average number of photons in a two-mode squeezed state is given by the square of the beta coefficient, this means that we would need around 100 repetitions in order to achieve successful single-photon detections. Putting everything together and considering a measurement time of around $1 \operatorname{\mu s}$ \cite{Ramsey} we can predict an event rate of approximately $kHz$ in the low photon number regime $n\simeq 3$, which could be improved with faster readout times, as in \cite{ibmqubits}.

It is important to remark that errors stemming from noisy state preparation, imperfect implementation of the unitary $U$ and measurement, will occur. This is one of the main limitations in any practical implementation of boson sampling, as the error rapidly scales with the size of the simulator. The implementation proposed in this work is not error-free either and, while the noise analysis may yield promising results as other circuit QED implementations of boson sampling \cite{Peropadre2015microwave}, a careful study to figure out error thresholds for scalability is still needed.

\subsection*{Complexity of Dynamical Casimir Effect}
Let us discuss the computational complexity inherent to a randomized DCE-like evolution. So far we have seen how DCE implemented in a two-coupled superconducting resonator system acts a quantum simulator of GBS by virtue of simple red- and blue-sideband dynamics. However, one could think of a more general scenario, in which the SQUID is fed with a multimode magnetic field 
\begin{equation}
\Phi_{\text{ext}}=\sum_{l,j=1}^{M}\Phi_{lj}^-\cos(\omega^{-}_{lj} t)+\Phi_{lj}^+\cos(\omega^{+}_{lj} t),
\end{equation}
implementing a Hamiltonian dynamics of of the form
\begin{equation}
H=\sum_{lj}^M a'^\dagger_l g_{lj} a_j + a'^\dagger_l \xi_{lj} a^\dagger_j + \text{H.c.},
\label{eq:GBS_Ham}
\end{equation}
which resembles a generalized boson sampling Hamiltonian \cite{PhysRevA.95.032327}. It is not difficult to realize that the generalized Anger-Jacobi expansion yields a one-to-one relation between the external field amplitudes $\Phi_{lj}^\pm$ and the coefficients $g_{lj},\xi_{lj}$ in terms of multivariate normal moments \cite{rahimi2015,Huh2014, rohde2016quantum,huh2016}. 
Using this mapping, and provided that the magnetic field amplitudes $\Phi_{lj}^\pm$ -more precisely the dimensionless ratio $\Phi^\pm_{lj}/\Phi_0$- are drawn from a random Haar measure, we conclude that a randomized DCE-like evolution lies outside the complexity class $\text{P}$, thus implementing a task that it is widely believed to be classically hard. 

While cutting edge signal generators are capable of creating train pulses with hundreds of frequencies, and resonators allocating hundreds of modes have been developed \cite{sundaresan2015beyond}, a sufficiently large randomized DCE experiment that exhibit quantum supremacy \cite{Peropadre2015microwave,preskill12} seems challenging due to the current limitation in resonator lifetimes and frequency-resolving measurements. A promising idea could be replace our standard transmission line resonators by left-handed transmission line  metamaterials, where a very dense mode spectrum has already been reported \cite{doi:10.1117/12.2180012}. However, we believe that this remarkable implication about the DCE computational complexity will trigger the forthcoming development of DCE-like experiments.

\section*{Discussion}
In summary, we have shown how to use the DCE in order to implement a GBS device. We propose a setup consisting of two superconducting 
transmission line resonators with different energy spectra that are coupled by a SQUID. The ultrafast modulation of the magnetic field fed into the SQUID by an external pump plays the role of relativistic motion of a mirror shared by the two resonators. The corresponding Bogoliubov transformation results into multimode parametric amplification. We show how a suitable choice of parameters allows to implement GBS and in particular scattershot BS, thus demonstrating that DCE can be used to implement a task that it is widely believed to be classically hard. Moreover, we show that randomized DCE-like dynamics should be itself classically hard.

\section*{Acknowledgments}
B.P. acknowledges the Air Force of Scientific Research for support under award: FA9550-12-1-0046. C.S. acknowledges financial support from Fundaci{\'o}n General CSIC (Programa ComFuturo). J. H. acknowledges supports by Basic Science Research Program through the National Research Foundation of Korea (NRF) funded by the Ministry of Education, Science and Technology (NRF-2015R1A6A3A04059773.

\section*{Competing interests}
The authors declare no conflict of interest

\section*{Author contributions}
B.P and C.S. conceived and designed the work. B.P.,J. H and C.S. worked on the theory, analyzed the data and wrote the manuscript. 

\end{document}